# High C/O Ratio and Weak Thermal Inversion in the Very Hot Atmosphere of Exoplanet WASP-12b

Nikku Madhusudhan[1], Joseph Harrington[2], Kevin B. Stevenson[2], Sarah Nymeyer[2], Christopher J. Campo[2], Peter J. Wheatley[3], Drake Deming[4], Jasmina Blecic[2], Ryan A. Hardy[2], Nate B. Lust[2], David R. Anderson[5], Andrew Collier-Cameron[6], Christopher B. T. Britt[2], William C. Bowman[2], Leslie Hebb[7], Coel Hellier[5], Pierre F. L. Maxted[5], Don Pollacco[8], Richard G. West[9]

*[1]Department of Astrophysical Sciences, Princeton University, Princeton, NJ 08544, USA (formerly at Massachusetts Institute of Technology, Cambridge, MA 02139, USA). [2]Planetary Sciences Group, Department of Physics, University of Central Florida, Orlando, FL 32816-2385, USA. [3]Department of Physics, University of Warwick, Coventry, CV4 7AL,UK. [4]NASA's Goddard Space Flight Center, Greenbelt, MD 20771-0001, USA. [5]Astrophysics Group, Keele University, Staffordshire ST5 5BG, UK. [6]School of Physics and Astronomy, University of St. Andrews, North Haugh, Fife KY16 9SS, UK [7]Department of Physics and Astronomy, Vanderbilt University, Nashville, TN 37235, USA. [8]Astrophysics Research Centre, School of Mathematics & Physics, Queen's University, University Road, Belfast, BT7 1NN, UK. [9]Department of Physics and Astronomy, University of Leicester, Leicester, LE1 7RH, UK*

**The carbon-to-oxygen ratio (C/O) in a planet provides critical information about its primordial origins and subsequent evolution. A primordial C/O greater than 0.8 causes a carbide-dominated interior as opposed to a silicate-dominated composition as found on Earth[1]; the solar C/O is 0.54 (Ref. 2). Theory states that high C/O leads to a diversity of carbon-rich planets that can have very different interiors and atmospheres from those in the solar system[1,3]. Here we report the detection of C/O ≥ 1 in a planetary atmosphere. The transiting hot Jupiter WASP-12b[4] has a dayside atmosphere depleted in water vapour and enhanced in methane**

**by over two orders of magnitude compared to a solar-abundance chemical equilibrium model at the expected temperatures. The observed concentrations of the prominent molecules[5,6] CO, $CH_4$, and $H_2O$ are consistent with theoretical expectations for an atmosphere with the observed C/O ≥ 1. If high C/O ratios are common, then some extrasolar planets are likely very different in interior composition from expectations based on solar abundances[1,3,7,8], and motivate new interior models attempting to explain the large diversity in observed radii. We also find that the extremely irradiated atmosphere (> 2500 K) of WASP-12b lacks a prominent thermal inversion, or a stratosphere, and has very efficient day-night energy circulation. The absence of a strong thermal inversion is in stark contrast to theoretical predictions for the most highly irradiated hot-Jupiter atmospheres[9,10,11].**

The transiting hot Jupiter WASP-12b orbits a star slightly hotter than the Sun (6300 K) in a circular orbit at a distance of only 0.023 AU, making it one of the hottest exoplanets known[4]. Thermal emission from the dayside atmosphere of WASP-12b has been reported using the Spitzer Space Telescope[12], at 3.6, 4.5, 5.8, and 8 μm wavelengths[13], and from ground-based observations in the J (1.2 μm), H (1.6 μm), and Ks (2.1 μm) bands[14] (Figure 1).

The observations provide constraints on the dayside atmospheric composition and thermal structure, based on the dominant opacity source in each bandpass. The J, H, and Ks channels[14] have limited molecular absorption features, and hence probe the deep layers of the planetary atmosphere, at pressure ($P$) of ~ 1 bar, where the temperature ($T$) is ~ 3000 K (Figure 1). The Spitzer observations[13], on the other hand, are excellent probes of molecular composition. $CH_4$ has strong absorption features in the 3.6 μm and 8 μm channels, CO has strong absorption in the 4.5 μm channel, and $H_2O$ has its strongest feature in the 5.8 μm channel and weaker features in the 3.6 μm, 4.5 μm, and 8 μm channels. The low brightness temperatures in the 3.6 μm (2700 K) and 4.5 μm (2500 K) channels, therefore, clearly suggest strong absorption due to $CH_4$ and CO, respectively. The high brightness temperature in the 5.8 μm channel, on the other hand, indicates low absorption due to $H_2O$. The strong CO absorption in the 4.5 μm channel also indicates temperature decreasing with altitude, since a thermal inversion would cause emission features of CO in the same channel with a significantly higher flux than at 3.6 μm[6,16].

The broadband observations allow us to infer the chemical composition and temperature structure of the dayside atmosphere of WASP-12b, using a statistical retrieval technique[6]. We combined a 1-D atmosphere model with a Markov-chain Monte Carlo sampler[6,17] that computes over $4\times10^6$ models to explore the parameter space. The phase space included thermal profiles with and without inversions, and equilibrium and non-equilibrium chemistry over a wide range of atomic abundances. Our models include the dominant sources of infrared opacity in the temperature regime of WASP-12b[5,18,19]: $H_2O$, CO, $CH_4$, $CO_2$, $H_2 – H_2$ collision induced absorption, and TiO and VO where the temperatures are high enough for them to exist in gas phase[9,20]. The host star has a significantly enhanced metallicity (2 x solar)[4], and evolutionary processes can further enhance the abundances[7,8]; Jupiter has 3 x solar C/H (Ref. 7). Our models therefore explore wide abundance ranges: 0.01 – 100 x solar for C/H and O/H, and 0.1 – 10 x solar for C/O. Figure 2 shows the mixing ratios of $H_2O$, CO, $CH_4$, and $CO_2$, and the ratios of C/H, O/H and C/O, required by the models at different levels of fit. Figure 3 presents the temperature profiles.

We find a surprising lack of water and overabundance of methane (Figure 2). At 2000 – 3000 K, assuming solar abundances yields CO and $H_2O$ as the dominant species besides $H_2$ and He[19,20]. Most of the carbon, and the same amount of oxygen, are present in CO, and some carbon exists as $CH_4$. The remaining oxygen in a hydrogen-dominated atmosphere is mostly in $H_2O$; small amounts are also present in species such as $CO_2$. The $CO/H_2$ and $H_2O/H_2$ mixing ratios should each be > 5 × $10^{-4}$, $CH_4/H_2$ should be < $10^{-8}$, and $CO_2/H_2$ should be ~ $10^{-8}$, under equilibrium conditions at a nominal pressure of 0.1 bar. The requirement of $H_2O/H_2 \leq 6 \times 10^{-6}$ and $CH_4/H_2 \geq 8 \times 10^{-6}$ (both at 3σ, 99.73% significance; Figure 2) is therefore inconsistent with equilibrium chemistry using solar abundances.

The observations place a strict constraint on the C/O ratio. We detect C/O ≥ 1 at 3σ significance (Figure 2). Our results rule out a solar C/O of 0.54 at 4.2σ. Our calculations of equilibrium chemistry[19, 29] with a C/O = 1 yield mixing ratios of $H_2O$, CO, and $CH_4$ that are consistent with the observed constraints. We find that, for C/O = 1, $H_2O$ mixing ratios as low as $10^{-7}$ and $CH_4$ mixing ratios as high as $10^{-5}$ can be attained in the 0.1 – 1 bar level for temperatures around 2000 K and higher. And, while the CO mixing ratio is predicted to be > $10^{-4}$, making it the dominant molecule after $H_2$ and He, $CO_2$ is predicted to be negligible (<$10^{-9}$). These theoretical predictions for a C/O = 1 atmosphere, are consistent with the observed constraints on $H_2O$, $CH_4$, CO, and $CO_2$ (Figure 2).

The observations rule out a strong thermal inversion deeper than 0.01 bar (Figure 3). Thermal inversions at lower pressures have opacities too low to induce features in the emission spectrum that current instruments can resolve. For comparison, all stratospheric inversions in solar system giant planets, and those consistent with hot Jupiter observations, exist at pressures between 0.01 – 1 bar[6,16,21]. The major contributions to all the observations come from the lower layers of the atmosphere, $P$ > 0.01 bar, where we rule out a thermal inversion (Figure 1 of SI). The observations also suggest very efficient day-night energy redistribution (Figure 2). The low brightness temperatures at 3.6 and 4.5 μm imply that only part of the incident stellar energy is reradiated from the dayside, while up to 45% is absorbed and redistributed to the nightside. The possibility of a deep thermal inversion and inefficient redistribution was suggested recently[14], based on observations in the J, H, and Ks channels, but the Spitzer observations rule out both conditions.

The lack of a prominent thermal inversion contrasts with recent work that designates WASP-12b to the class of very hot Jupiters that are expected to host inversions[9,22]. At $T$ > 2000K, molecules such as TiO and VO, which are strong absorbers in the UV/visible, are expected to be in gas phase and potentially cause thermal inversions[9]. WASP-12b, now being the hottest planet without a distinct inversion, presents a major challenge to existing atmospheric classification schemes for exoplanets based on thermal inversions[9,22]. Although there are hints of low chromospheric activity[10] in the host star, it remains to be seen if the high incident continuum UV flux expected for WASP-12b might be efficient in photo-dissociating inversion-causing compounds, thus explaining the lack of a strong inversion[10]. Alternatively, the amount of vertical mixing might be insufficient to keep TiO/VO aloft in the atmosphere to cause thermal inversions[20]. A C/O = 1 might also yield lower TiO/VO than that required to cause a thermal inversion. It is unlikely that TiO/VO in WASP-12b might be lost to cold traps[20], given the high temperatures in the deep atmosphere on the day and night sides.

If high C/O ratios are common, then the formation processes and compositions of extrasolar planets are likely very different from expectations based on solar system planets. The host star has super-solar metallicity but initial analyses find its C/O consistent with solar[4,23]. In the core accretion model, favoured for the formation of Jupiter, icy planetesimals containing heavy elements coalesce to form the core, followed by gas accretion[8,24]. The abundances of elemental oxygen and carbon are enhanced equally[7,8], maintaining a C/O like the star's. If the host star had a C/O ~ 1, then the C/O we detect in WASP-12b would have been evident. However, if the stellar C/O is indeed

< 1, then the C/O enhancement in WASP-12b's atmosphere would suggest either an unexpected origin for the planetesimals, a local over-density of carbonaceous grains[3,25], or a different formation mechanism entirely. Although carbon-rich giant planets like WASP-12b have not been observed, theory predicts myriad compositions for carbon-dominated solid planets[1,3]. Terrestrial-sized carbon planets, for instance, could be dominated by graphite or diamond interiors, as opposed to the silicate composition of Earth[1,3]. If carbon dominates the heavy elements in the interior of a hot Jupiter, estimates of mass and radius could change compared to those based on solar abundances. Future interior models[26] should investigate the contribution of high C/O to the large radius of WASP-12b: 1.75 Jupiter radii for 1.4 Jupiter masses (Ref. 4).

The observed molecular abundances in the dayside atmosphere of WASP-12b motivate the exploration of a new regime in atmospheric chemistry. It remains to be seen if photochemistry in WASP-12b can significantly alter the composition in the lower layers of the atmosphere, $P = 0.01 - 1$ bar, which contribute most to the observed spectrum (Figure 1 of SI). Explaining the observed composition as a result of photochemistry with solar abundances would be challenging. $CH_4$ is more readily photodissociated than $H_2O$[11,27], and hence a depletion of $CH_4$ over that predicted with solar abundances might be expected, as opposed to the observed enhancement of $CH_4$. Apart from the spectroscopically dominant molecules considered in this work, minor species such as OH, $C_2H_2$, and FeH (Refs. 27, 28), which are not detectable by current observations, could potentially be measured with high-resolution spectroscopy in the future. Detection of these species would allow additional constraints on equilibrium and non-equilibrium chemistry in WASP-12b, although their effect on the C/O would be negligible. Models of exoplanetary atmospheres have typically assumed solar abundances and/or solar C/O, thereby exploring a very limited region of parameter space[9,16,29]. Data sufficient for a meaningful constraint on C/O exist for only a few exoplanets. That this initial C/O statistical analysis has C/O ≥ 1 potentially indicates a wide diversity of planetary compositions.

1. Bond, J. C., O'Brien, D. P., & Lauretta, D. S. The compositional diversity of extrasolar planets. I. in situ simulations. *The Astrophysical Journal*, **715**, 1050-1070 (2010).

2. Asplund, M., Grevesse, N., & Sauval, A., The solar chemical composition. ASP Conf. Ser. **336**, Cosmic Abundances as Records of Stellar Evolution and Nucleosynthesis, ed. T. G. Barnes, III & F. N. Bash (San Francisco, CA: ASP), 25-38 (2005)

3. Kuchner, M. & Seager, S. Extrasolar carbon planets. arXiv:astro-ph/0504214 (2005)

4. Hebb, L. et al. 2009, WASP-12b: The Hottest Transiting Extrasolar Planet Yet Discovered. *The Astrophysical Journal*, **693**, 1920-1928 (2009)

5. Swain, M. R. *et al.* Molecular Signatures in the Near-Infrared Dayside Spectrum of HD 189733b. *Astrophysical Journal Letters,* **690**, L114–L117 (2009).

6. Madhusudhan, N. & Seager, S. A temperature and abundance retrieval method for exoplanet atmospheres, *The Astrophysical Journal*, **707**, 24-39 (2009).

7. Atreya, S. K. & Wong, A. S. Coupled clouds and chemistry of the giant planets – A case for multiprobes. *Space Science Reviews*, 116, 121-136 (2005)

8. Owen, T. et al. A low-temperature origin for the planetesimals that formed Jupiter. *Nature*, **402**, 269-270 (1999)

9. Fortney, J. J., Lodders, K., Marley, M. S. & Freedman, R. S., A Unified Theory for the Atmospheres of the Hot and Very Hot Jupiters: Two Classes of Irradiated Atmospheres. *The Astrophysical Journal*, **678**, 1419-1435 (2008)

10. Knutson, H. A., Howard, A. W. & Isaacson, H. A Correlation Between Stellar Activity and Hot Jupiter Emission Spectra. *The Astrophysical Journal*, 720, 1569 – 1576 (2010)

11. Zahnle, K. et al. Atmospheric Sulfur Photochemistry on Hot Jupiters. *The Astrophysical Journal*, 701, L20-L24 (2009).

12. Werner, M. W. *et al.* The Spitzer Space Telescope Mission. *Astrophy. J. Suppl. Ser.* **154**, 1–9 (2004).

13. Campo, C. et al. On the Orbit of Exoplanet WASP-12b. arXiv:astro-ph/1003.2763 (2010)

14. Croll, B. et al. Near-infrared thermal emission from WASP-12b: detections of the secondary eclipse in Ks, H & J. arXiv:astro-ph/1009.0071v2 (2010)

15. Lopez-Morales et al. Day-side z'-band Emission and Eccentricity of WASP-12b. *The Astrophysical Journal*, **716**, L36-L40 (2010)

16. Burrows, A., Budaj, J., & Hubeny, I. Theoretical Spectra and Light Curves of Close-in Extrasolar Giant Planets and Comparison with Data. *The Astrophysical Journal*, **678**, 1436-1457 (2008)

17. Gilks, W.R., Richardson, S., & Spiegelhalter, D. J. *Markov Chain Monte Carlo in Practice*, Chapman & Hall, London (1996)

18. Lodders, K. & Fegley, B. Atmospheric Chemistry in Giant Planets, Brown Dwarfs, and Low-Mass Dwarf Stars. I. Carbon, Nitrogen, and Oxygen. *Icarus* **155**, 393-424 (2002)

19. Burrows, A. & Sharp, C. M. Chemical Equilibrium Abundances in Brown Dwarf and Extrasolar Giant Planet Atmospheres. *The Astrophysical Journal,* **512**, 843–863 (1999).

20. Spiegel, D. S., Silverio, K., & Burrows, A., Can TiO Explain Thermal Inversions in the Upper Atmospheres of Irradiated Giant Planets? *The Astrophysical Journal*, **699**, 1487-1500 (2009)

21. Yung, Y. & DeMore, W. B. Photochemistry of Planetary Atmospheres. *Oxford University Press, New York.* (1999)

22. Hubeny, I., Burrows, A., \& Sudarsky, D., A possible bifurcation in atmospheres of strongly irradiated stars and planets. *The Astrophysical Journal*, **594**, 1011-1018 (2003)

23. Fossati, L. et al. A detailed spectropolarimetric analysis of the planet-hosting star WASP-12, *The Astrophysical Journal*, **720**, 872-886 (2010)

24. Pollack, J. B. et al. Formation of the Giant Planets by Concurrent Accretion of Solids and Gas. *Icarus*, **124**, 62-85 (1996)

25. Lodders, K. Jupiter formed with more tar than ice. *The Astrophysical Journal*, **611**, 587-597 (2004)

26. Fortney, J. J., Marley, M. S., \& Barnes, J. W., Planetary radii across five orders of magnitude in mass and stellar insolation: application to transits, *The Astrophysical Journal*, **659**, 1661-1672 (2007)

27. Line, M. R., Liang, M. C., & Yung, Y. L. High-temperature Photochemistry in the Atmosphere of HD 189733b. *The Astrophysical Journal*, 717, 496-502 (2010).

28. Cushing, M. C., Rayner, J. T., & Vacca, W. D. An Infrared Spectroscopic Sequence of M, L, and T Dwarfs. *The Astrophysical Journal*, **623**, 1115-1140 (2005)

29. Seager, S. et al. On the Dayside Thermal Emission of Hot Jupiters. *The Astrophysical Journal*, **632**, 1122-1131 (2005)

30. Castelli, F. & Kurucz, R. L. New Grids of ATLAS9 Model Atmospheres. *ArXiv Astrophysics e-prints* (2004). arXiv:astro-ph/0405087.

**Acknowledgements:** We thank the authors of Ref 14 for sharing their ground-based observations before publication, and Thomas J. Loredo for helpful discussions. NM thanks Sara Seager for financial support during his stay at MIT where most of the modelling work was carried out. This work is based on observations made with the Spitzer Space Telescope, which is operated by the Jet Propulsion Laboratory, California Institute of Technology under a contract with NASA. Support for this work was provided by NASA through an award issued by JPL/Caltech.

**Contributions:** NM conducted the atmospheric modelling and wrote the paper with input on both from JH. JH and PJW led the observing proposals, data from which have been interpreted in this work. JH, JB, and CJC designed the observations with input from PJW, DRA, AC-C, LH, CH, PFLM, DP, and RGW. JH, KBS, SN, CJC, DD, JB, RAH, NBL, DRA, AC-C, CBTB, and WCB analyzed the Spitzer data.

The authors declare no competing financial interests.

**Author Information:** Correspondence and requests for materials should be addressed to N.M. (e-mail: nmadhu@astro.princeton.edu, nmadhu@mit.edu).

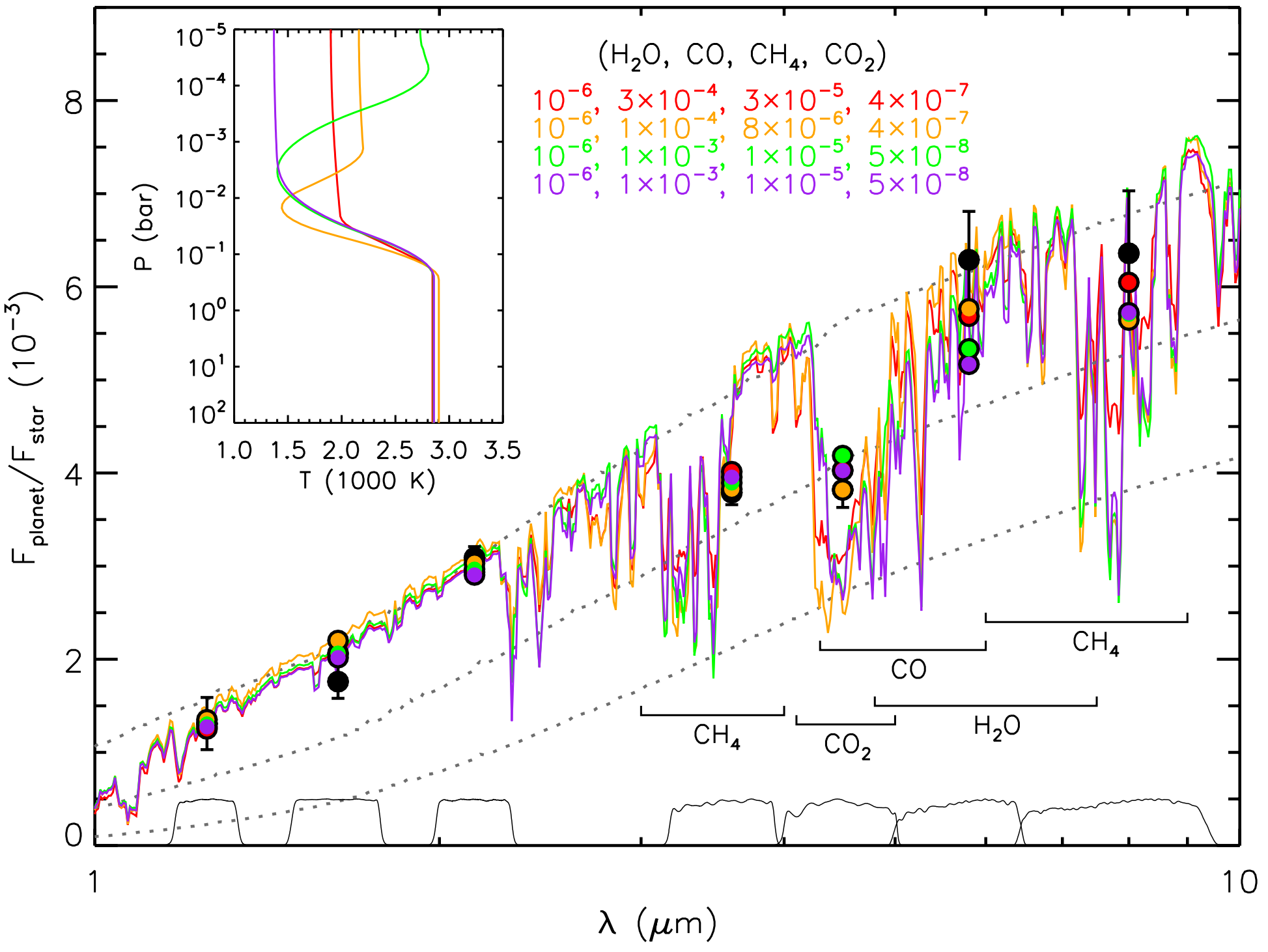

Figure 1: Observations and model spectra for dayside thermal emission of WASP-12b. The black filled circles with error bars show the data: four Spitzer observations[13] (3.6 μm, 4.5 μm, 5.8 μm, and 8 μm), and three ground-based observations in the J (1.2 μm), H (1.6 μm), and Ks (2.1 μm) bands[14]. Four models fitting the observations are shown in the coloured solid curves in the main panel, and the coloured circles are the channel-integrated model points. The corresponding temperature profiles are shown in the inset. The molecular compositions are shown as number ratio with respect to molecular hydrogen; all the models have C/O between 1-1.1. The thin gray dotted lines show blackbody spectra of WASP-12b at 2000 K (bottom), 2500 K, and 3000 K (top). A Kurucz model[30] was used for the stellar spectrum, assuming uniform illumination over the planetary disk (i.e., weighted by 0.5; Ref 10). The black solid lines at the bottom show the photometric band-passes in arbitrary units. The low fluxes at 3.6 and 4.5 μm are explained by methane and CO absorption, respectively, required for all the models that fit. The high flux in the 5.8 μm channel indicates

less absorption due to $H_2O$. The observations can be explained to high precision by models without thermal inversions. Models with strong thermal inversions are ruled out by the data (see Figure 3). The green model features a thermal inversion at low pressures ($P$ < 0.01 bar), but the corresponding spectrum is almost indistinguishable from the purple model, which has identical composition and thermal profile below the 0.01 bar level to the green model, but does not have a thermal inversion above the 0.01 bar level. Thus, any potential thermal inversion is too weak to be detectable by current instruments.

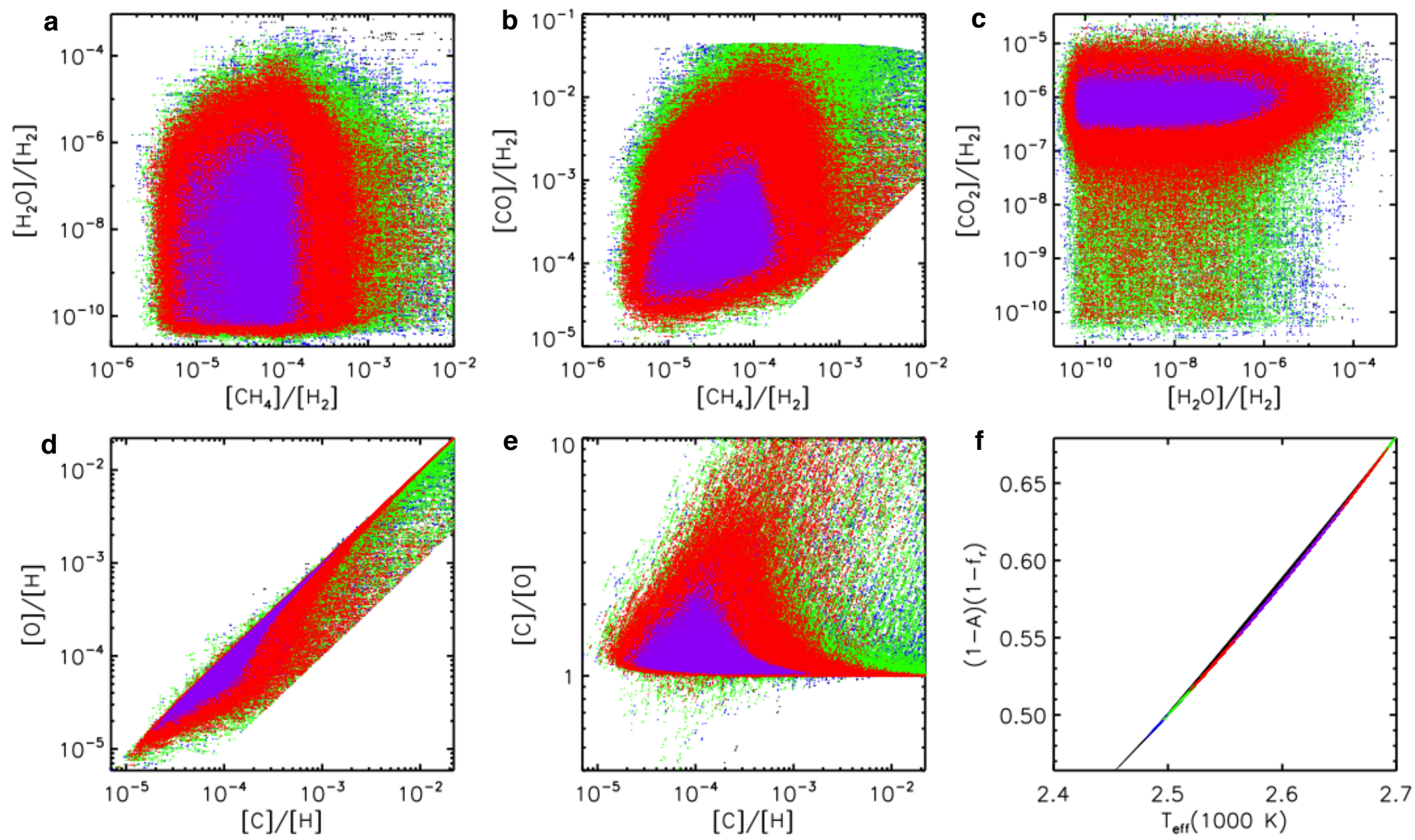

Figure 2: Constraints on the atmospheric composition of WASP-12b. The distributions of models fitting the 7 observations (Figure 1) at different levels of $\chi^2$ are shown. The coloured dots show $\chi^2$ surfaces, with each dot representing a model realization. The purple, red, green, blue, and black colours correspond to models with $\chi^2$ less than 7, 14, 21, 28, and $\chi^2$ > 28, respectively ($\chi^2$ ranges between 4.8 – 51.3). Mixing ratios are shown as ratios by number with respect

to $H_2$. At 3$\sigma$ significance, the constraints on the composition are $H_2O/H_2 \leq 6 \times 10^{-6}$, $CH_4/H_2 \geq 8 \times 10^{-6}$, $CO/H_2 \geq 6 \times 10^{-5}$, $CO_2/H_2 \leq 5 \times 10^{-6}$, and C/O > 1. The compositions of the best-fitting models (with $\chi^2 < 7$) span $H_2O/H_2 = 5 \times 10^{-11} – 6 \times 10^{-6}$, $CO/H_2 = 3 \times 10^{-5} – 3 \times 10^{-3}$, $CH_4/H_2 = 4 \times 10^{-6} – 8 \times 10^{-4}$, and $CO_2/H_2 = 2 \times 10^{-7} – 7 \times 10^{-6}$; the corresponding ranges in C/O and elemental abundances are C/O = 1 – 6.6, C/H = $2 \times 10^{-5} – 10^{-3}$ and O/H = $2 \times 10^{-5} – 10^{-3}$. The constraints on the C/H and O/H ratios are governed primarily by the constraints on CO, which is the dominant molecule after $H_2$ and He. Based on thermo-chemical equilibrium, the inferred $CH_4/H_2$ and $H_2O/H_2$ mixing ratios are possible only for C/O ≥ 1, consistent with our detection of C/O ≥ 1. The last panel shows the constraints on the day-night energy redistribution[6], given by $(1-A)(1-f_r)$, where $A$ is the bond albedo and $f_r$ is the fraction of incident energy redistributed to the night side. Up to $f_r = 0.45$ is possible (for $A = 0$). Thus, the observations support very efficient redistribution. An additional observation in the z' (0.9 µm) band was reported recently[15]. However, the observation implies a value for the orbital eccentricity inconsistent with other data in the literature[13, 14]. We therefore decided to exclude this observation from the analysis presented here, although including it does not affect our conclusions regarding the value of C/O or the temperature structure.

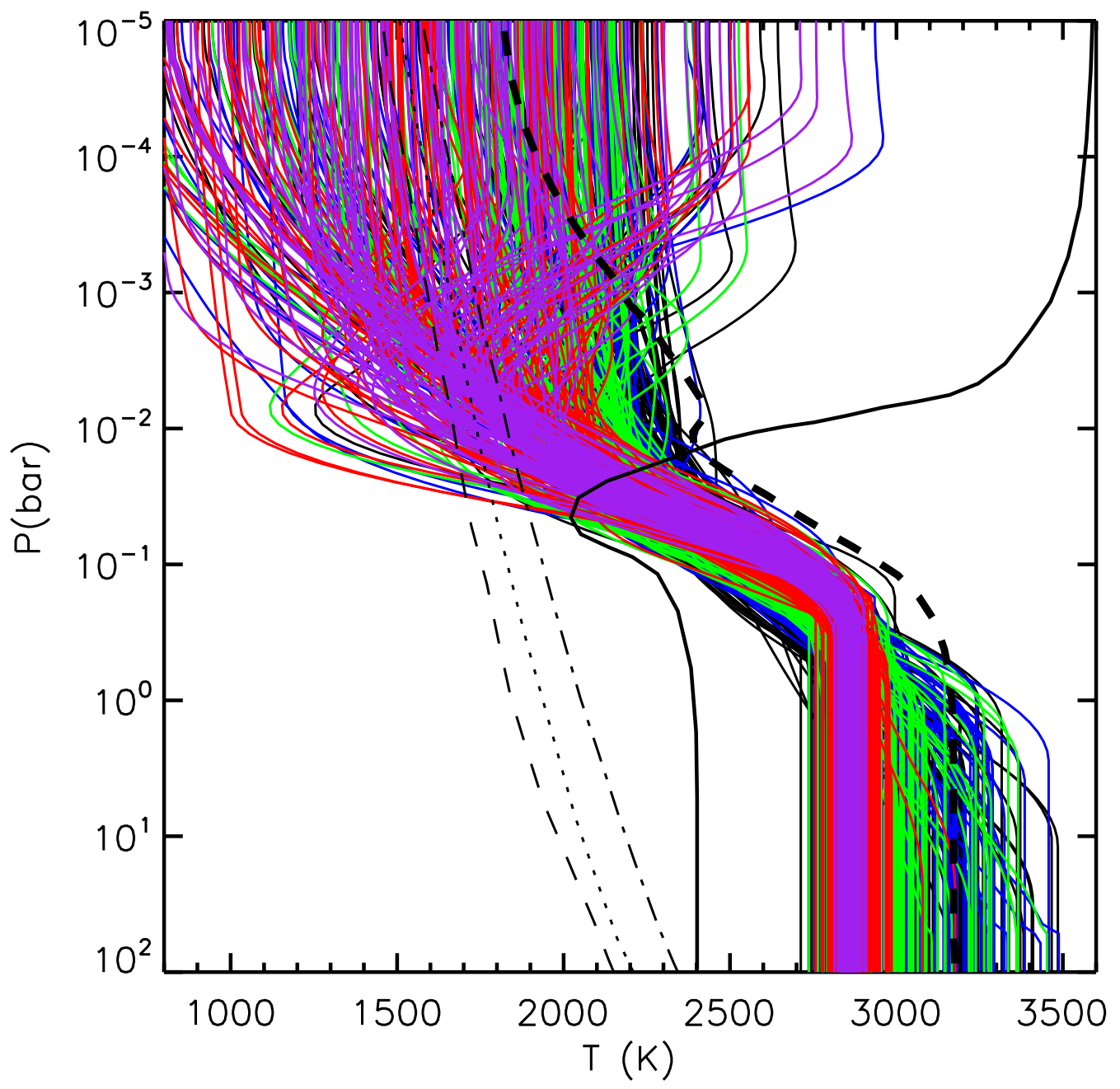

Figure 3: Thermal profiles of WASP-12b. The solid thin lines show profiles at different degrees of fit (description of colours is same as in Figure 2); only 100 randomly chosen profiles for each $\chi^2$ level are shown, for clarity. The thick, black, solid (dashed) curve in the front shows a published profile from a self-consistent model of WASP-12b with (without) a thermal inversion, adapted from Ref. 20, which assumes solar abundances. If a thermal inversion is present in WASP-12b, it is expected to be prominent, as shown by the thick solid black curve. A prominent thermal inversion between 0.01 – 1 bar is ruled out by the data at $4\sigma$. The ostensibly large inversions in the figure are at low pressures (below 0.01 bar), which have low optical depths, and hence minimal influence on the emergent spectrum (see Figure 1). The observations are completely consistent with thermal profiles having no inversions. Small thermal inversions are also admissible by the data, and could potentially result from dynamics. The thin dotted, dashed, and dash-dot lines in black show condensation curves of TiO at solar, $0.1 \times$ solar, and $10 \times$ solar composition[20].

# Supplementary Information

## A. Contribution Functions

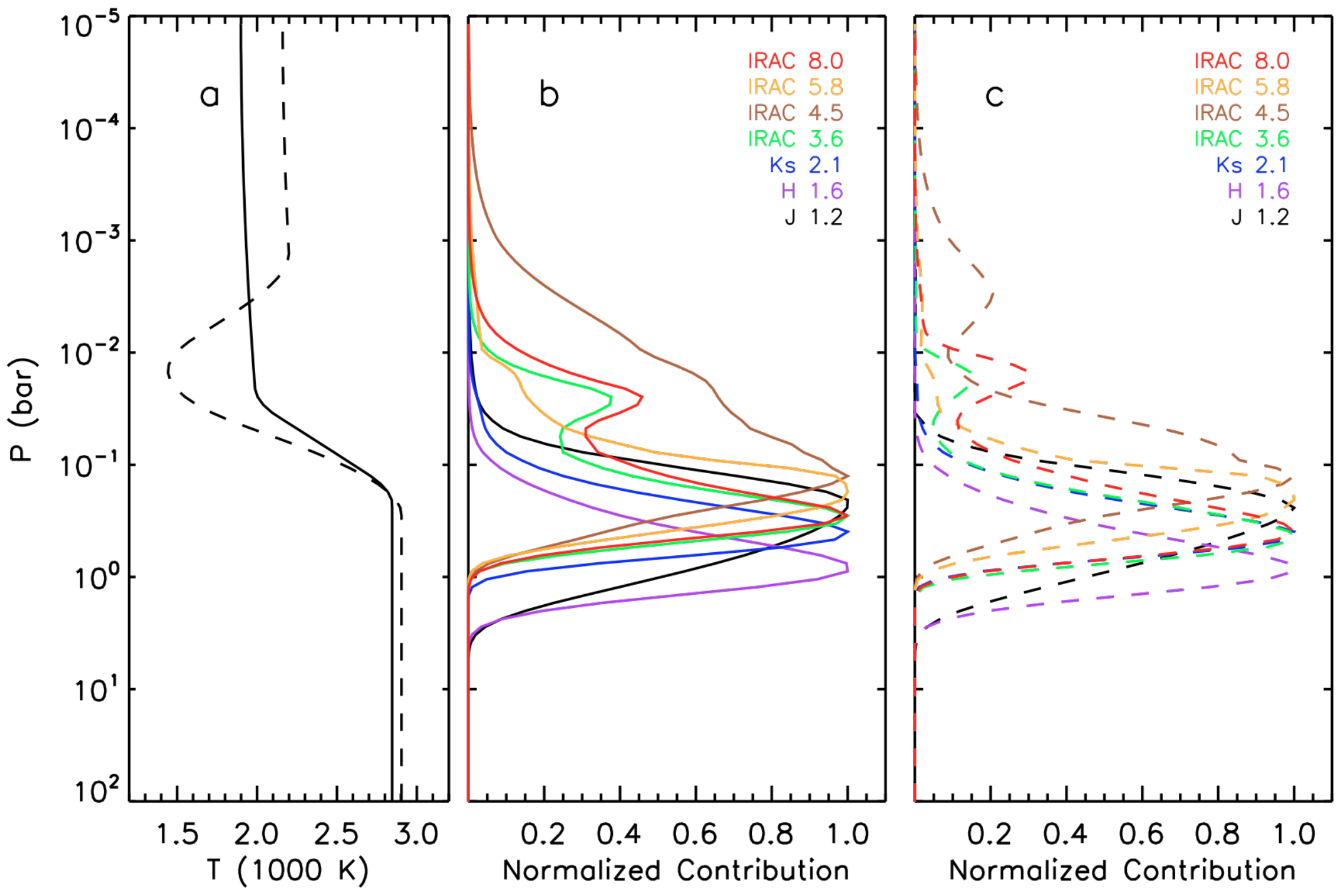

SI Figure 1: Contribution functions for representative WASP-12b models showing the atmospheric origin of flux observed in each bandpass. Two representative temperature profiles are shown in panel a (same as the red and orange profiles of Figure 1). The contribution functions in panel b (panel c) correspond to the solid (dashed) temperature profile in panel a, colour-coded by bandpass. The maximum contribution to the emergent flux of WASP-12b in all the channels comes from the lowest layers of the observable atmosphere, below the 0.01 bar level.

## B. Atmospheric Model and Parameter Space Exploration

We use the Markov-chain Monte Carlo (MCMC) method to explore the model parameter space. The MCMC method is a Bayesian parameter estimation algorithm, which allows the calculation of posterior probability distributions of model parameters conditional to a given set of observations (Ref 17). The MCMC method efficiently explores the parameter space near a global solution, subject to prior knowledge (e.g., of viable parameter ranges), to generate a posterior distribution. In this work, the number of observations ($N_{obs}$ = 7) is less than the number of model parameters ($N_{par}$ = 10), rendering the problem under-constrained, with no unique solution. However, it is still possible to explore the parameter space and determine contours in the error surface using a Bayesian approach.

We use the MCMC method with a Metropolis-Hastings scheme to sample the parameter space of a 1-D dayside atmosphere model of WASP-12b. The model follows from Ref. 6, and uses the same parameterization, assuming uniform priors. An additional constraint on the model is imposed in the form of energy balance, i.e., the integrated emergent energy from the planet must be lower than, or equal to, the incident stellar energy. A Kurucz model[30] was used for the stellar spectrum, assuming uniform illumination over the planetary disk (i.e., weighted by 0.5; Ref 10). We compute four chains of $10^6$ links each, with different initial conditions, and spanning models with and without thermal inversions. For a given set of parameters at each step of a chain, the $\chi^2$ statistic is evaluated as:

$$\chi^2 = \sum_{i=1}^{N_{obs}} \left( \frac{f_{i,m} - f_{i,obs}}{\sigma_i^{obs}} \right)^2 ,$$

where, $f_{i,m}$ and $f_{i,obs}$ are the modelled and observed planet-star flux contrasts, respectively, at each of the observed wavelengths, and $\sigma_{i,obs}$ is the corresponding observational uncertainty. All the chains lead to the same conclusions. Figure 2 and Figure 3 show the phase space spanned by all the chains, with each model realization colour-coded by $\chi^2$.

In Figure 2, the $\chi^2$ in the space of atmospheric composition is a nominal measure of model fit to data. For instance, one can consider models with $\chi^2 < 7$ as well fit models. However, the $\chi^2$ cannot be directly interpreted as a confidence measure because the number of degrees of freedom is negative. We estimate the statistical significances of the detected chemical composition by integrating over the posterior probability distributions of the required parameters. The results are presented in the paper. The under-constrained model cannot uniquely identify the specific C/O ratio in WASP-12b's atmosphere. However, the Bayesian integral can demonstrate that the entire region of phase space where C/O < 1 is ruled out at the 3-sigma level, and that where C/O ≤ 0.54 (solar value) is ruled out at 4.2-sigma.